\documentclass[a4paper,preprint,aip,apl,tightenlines,notitlepage]{revtex4-1}
\usepackage[utf8]{inputenc}
\usepackage{mathptmx}
\usepackage[symbol]{footmisc}
\usepackage[symbol]{footmisc}
\usepackage{graphicx}
\usepackage{subcaption}
\usepackage{textcomp}
\usepackage{amsmath}
\usepackage[hidelinks]{hyperref}

\begin{document}

\title{A tuneable telecom-wavelength entangled light emitting diode}

\author{Z.-H. Xiang}
\affiliation{Toshiba Research Europe Limited, Cambridge Research Laboratory, 208 Science Park, Milton Road, Cambridge, CB4 0GZ, UK}
\affiliation{Cavendish Laboratory, University of Cambridge, J.J. Thompson Avenue, Cambridge, CB3 0HD, UK}

\author{J. Huwer}
\email{jan.huwer@crl.toshiba.co.uk}

\author{J. Skiba-Szymanska}

\author{R. M. Stevenson}

\author{D. J. P. Ellis}
\affiliation{Toshiba Research Europe Limited, Cambridge Research Laboratory, 208 Science Park, Milton Road, Cambridge, CB4 0GZ, UK}

\author{I. Farrer}
\altaffiliation[Present address:]{Department of Electronic \& Electrical Engineering, University of Sheffield, Sheffield, S1 3JD, UK}
\affiliation{Cavendish Laboratory, University of Cambridge, J.J. Thompson Avenue, Cambridge, CB3 0HD, UK}

\author{M. B. Ward}
\affiliation{Toshiba Research Europe Limited, Cambridge Research Laboratory, 208 Science Park, Milton Road, Cambridge, CB4 0GZ, UK}

\author{D. A. Ritchie}
\affiliation{Cavendish Laboratory, University of Cambridge, J.J. Thompson Avenue, Cambridge, CB3 0HD, UK}

\author{A. J. Shields}
\affiliation{Toshiba Research Europe Limited, Cambridge Research Laboratory, 208 Science Park, Milton Road, Cambridge, CB4 0GZ, UK}

\begin{abstract}
Entangled light emitting diodes based on semiconductor quantum dots are promising devices for security sensitive quantum network applications, thanks to their natural lack of multi photon-pair generation. Apart from telecom wavelength emission, network integrability of these sources ideally requires electrical operation for deployment in compact systems in the field. For multiplexing of entangled photons with classical data traffic, emission in the telecom O-band and tuneability to the nearest wavelength channel in compliance with coarse wavelength division multiplexing standards (20 nm channel spacing) is highly desirable. Here we show the first fully electrically operated telecom entangled light emitting diode with wavelength tuneability of more than 25nm, deployed in an installed fiber network. With the source tuned to 1310.00 nm, we demonstrate multiplexing of true single entangled photons with classical data traffic and achieve entanglement fidelities above 95\% on an installed fiber in a city.
\end{abstract}

\maketitle

The success or failure of quantum light sources in advanced commercial photonic quantum-network applications is strongly dependent on integrability of these sources with current infrastructure and technology. Regarding networks, operation at telecom wavelength is as essential as wavelength tuneability, required for interfacing and multiplexing with other sources of classical or quantum light over the same optical fiber. In terms of quantum light emitters, scalable manufacturing techniques for their production as well as compliance with low voltage driving electronics for safe and long-term reliable operation in remote non-laboratory environments are most desirable. 

Good suppression of multi-photon emission is one of the cornerstones for the next level of high-speed quantum network applications, going beyond conventional quantum key distribution (QKD) \cite{shor2000simple} that is currently making use of weak coherent laser pulses \cite{lo2005decoy, lo2012measurement, sasaki2011field} or spontaneously generated entangled photon pairs \cite{curty2004entanglement, ursin2007entanglement, yin2017satellite}. Single photon-pair output paired with high fidelity entanglement will enable advanced communication modules based on quantum teleportation \cite{kimble2008quantum, cirac1997quantum} that are expected to deliver unconditional security \cite{gisin2002quantum, scarani2009security} and longer transmission distances \cite{pirandola2017fundamental}.

Quantum light sources based on III-V compound semiconductor quantum dots (QD) embedded in a p-i-n diode \cite{benson2000regulated} are considered a promising approach in terms of scalable manufacturability as they are sharing the same material platform as standard laser diodes. These devices generate single entangled pairs of photons on the so-called biexciton cascade after electrical injection of carriers, which is why they are also referred to as entangled light emitting diodes (ELED) \cite{salter2010entangled}. For quantum communication applications, careful engineering of the material composition and wafer growth conditions have enabled fabrication of ELEDs emitting in the standard telecom wavelength bands \cite{zinoni2006time, ward2014coherent, muller2018quantum}.

True network integration of quantum light sources requires co-existence with classical data traffic mostly in the telecom C-band over the same optical fiber. This favours operation of ELEDs in the telecom O-band as the large spectral separation combined with the generally reduced Raman scattering from the lower energy C-band to the higher energy O-band makes wavelength division multiplexing of quantum light at the single photon level with classical communication signals feasible \cite{chapuran2009optical, choi2011quantum, ciurana2014quantum}. More sophisticated network architectures requiring higher bandwidth make use of the so-called Coarse Wavelength Division Multiplexing (CWDM) grid (ITU G.694.2), comprising a total of 18 channels from 1271nm to 1611nm with 20nm spacing. For compatibility of ELEDs with this standard, spectral tuning ranges of 20nm are desirable such that any emitter can be tuned to the closest CWDM channel or even changed to a different channel. Despite there has been good progress in the deterministic growth of QDs emitting at specific non-telecom wavelengths \cite{jahn2015artificial, keil2017solid}, post-growth tuning mechanisms remain most important for enhancing the yield in manufacturability of devices. 

The application of magnetic fields \cite{schulhauser2002magneto} or strain \cite{zhang2015high, bonato2011strain, chen2016wavelength, jons2011dependence, zeuner2018stable} has been demonstrated for tuning QD emission. While achieving only a few nanometers of tuneability, these approaches require large magnetic fields or very high voltages which is not practical for network deployment. In contrast, wavelength tuning via static electric fields  \cite{patel2010two, oulton2002manipulation, findeis2001photocurrent} is suitable for long-term safe operation, achieving good tuning ranges with only low applied voltages. In this work, we demonstrate the first tuneable telecom ELED with low voltage operation, large tuning range and strong single-photon character, which is compatible with networks containing classical communication traffic. This enables us for the first time to operate an ELED outside of a research laboratory and show sub-Poissonian entangled photon transmission with high fidelity multiplexed with classical data signals over the metropolitan fiber network in Cambridge. 

\begin{figure}[!ht]
    \includegraphics[width=0.87\textwidth]{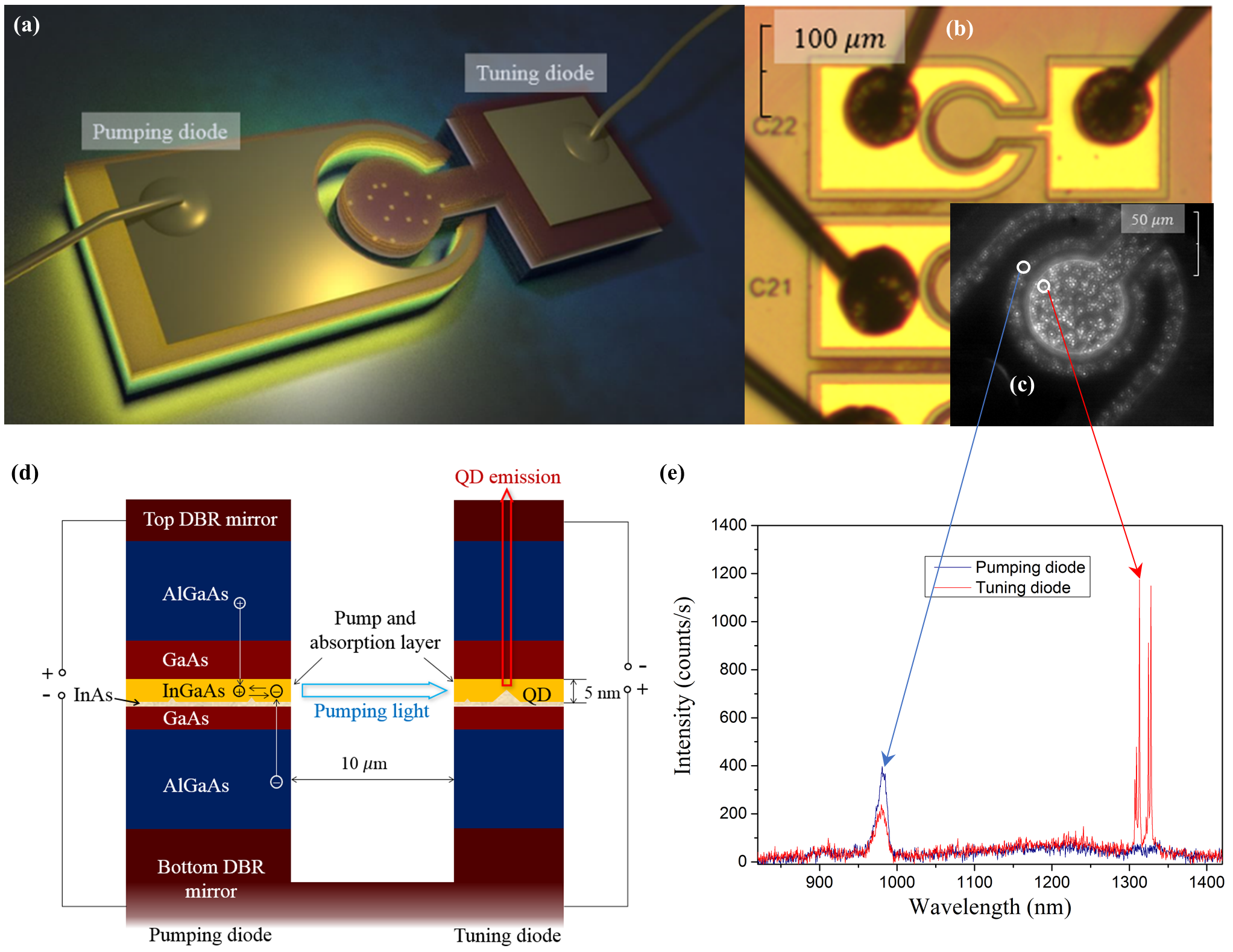}
  \caption{(a) Concept of tuneable entangled light emitting diode. A current is injected via the top contact on the pumping diode for optical excitation of single quantum dots in the embraced circular tuning diode. Both diodes share the same bottom contact not shown in the figure. (b) Microscope image of fabricated devices with attached bond wires. (c) Microscope image of QDs in the structure emitting at telecom wavelength when applying a forward bias to the pumping diode. A long-pass filter at 1100nm is placed before the camera to suppress short-wavelength pumping light for better visibility of QD emission. (d) Layer structure of wafer. The central 5nm InGaAs layer serves as pump and absorption layer for the on-chip optically excited device. It generates light at 950-1000nm in the pumping diode (left) which is absorbed in the tuning diode (right), exciting QDs. The shown layer thicknesses are not to scale. For more details see text. (e) Comparison of the spectrum of micro electro luminescence on the pumping diode (blue) and resulting micro photo luminescence from the tuning diode (red). The measurements were taken in a confocal microscope setup with collection from the top and a spot diameter of 2\(\mu m\).}
\label{fig:OCP}
\end{figure}

A schematic of the tuneable device design is shown in Figure \ref{fig:OCP}(a). Entangled photon pairs are emitted from the QDs located at the central circular region. These are optically excited by light emitted from the so-called ‘pumping diode’ surrounding it, when applying a forward bias to the top contact on the left. A reverse bias applied to the top contact from the inner diode is used to apply static fields for tuning the QD emission, making this the so-called ‘tuning diode’. The ring-circle structure enables homogeneous illumination of the tuning diode with pumping light from all sides. An optical microscope image of fabricated devices is shown in Figure \ref{fig:OCP}(b). The central mesa for the design used in this work has a radius of 35\(\mu m\), typically containing around 120 QDs as can be seen in the camera image in Figure \ref{fig:OCP}(c), showing luminescence of the QDs during operation of the device. 

The key features of the wafer structure are illustrated in  Figure \ref{fig:OCP}(d). The InAs/GaAs quantum dots used in this work were grown using the Stranski-Krastanow (S-K) growth mode in Molecular Beam Epitaxy. Use of the so-called bi-modal growth mode \cite{ward2014coherent} results in QD emission in the telecom O-band. The QD layer is deposited on GaAs and capped with a thick layer of InGaAs (5nm) which serves as optical pump and absorption layer, in contrast to structures operating at non-telecom wavelengths \cite{lee2017electrically, munnelly2017electrically}. The optically active part of the wafer is finished with another layer of GaAs, forming a quantum well (QW). Additional barriers made from AlGaAs grown below and above these central layers inhibit escape of charge carriers when applying electric fields along the vertical direction. The structure further contains doped stacked DBR mirrors made from GaAs/AlGaAs at the bottom and top, forming a p-i-n junction in a weak vertical \(\lambda/2\) cavity for enhancing the emission in the telecom O-band.

\begin{figure}[!t]
\centering
\begin{subfigure}{0.7\textwidth}
                \includegraphics[width=\textwidth]{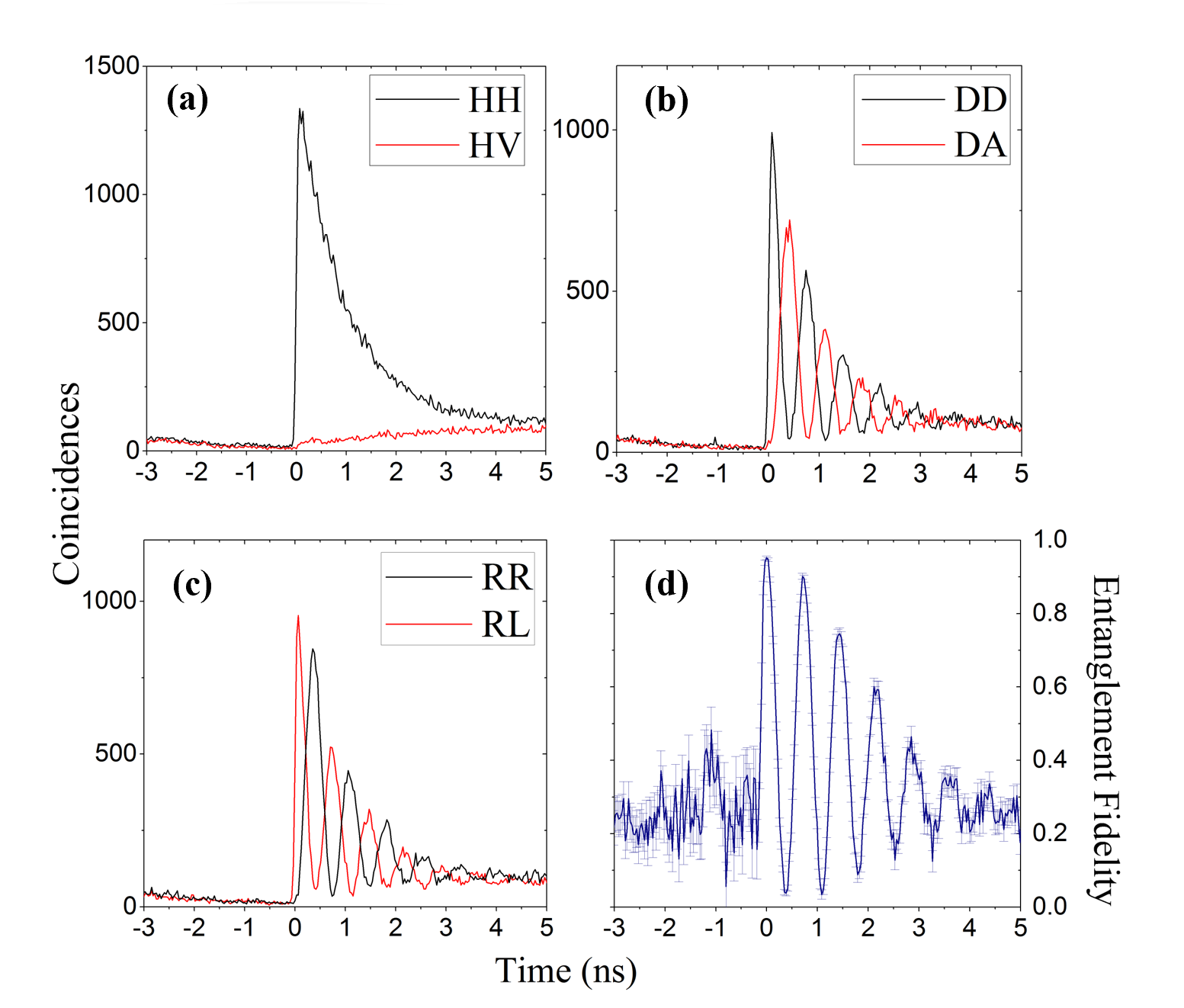}
        \end{subfigure}%

        \begin{subfigure}{0.7\textwidth}
                \includegraphics[width=\textwidth]{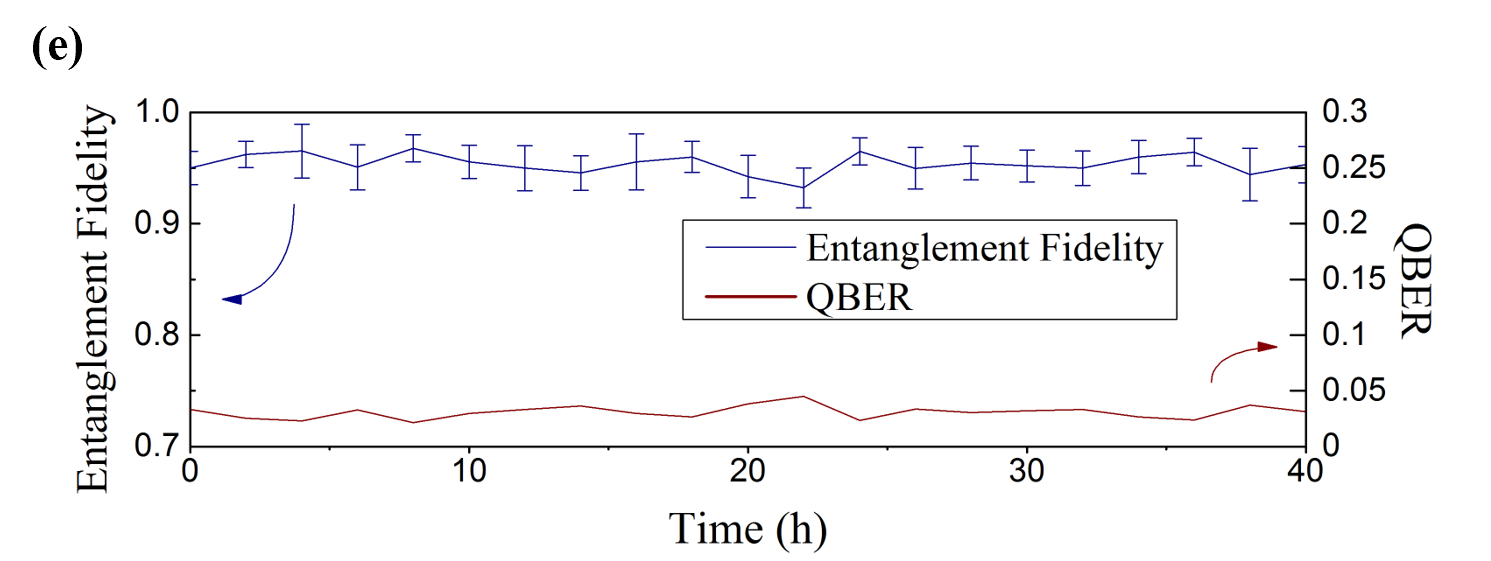}
        \end{subfigure}%
        \caption{Correlation measurements of entangled photons across installed network link. (a) HV basis; (b) DA basis; (c) RL basis. (d) Fidelity to Bell \(\phi^+\) state. Data is displayed on 48ps resolution timing grid. (e) Evolution of maximum fidelity and corresponding quantum bit error rate (QBER) over 40h of continuous operation. Each data point corresponds to 2h of data.}
    \label{fig:cor}
\end{figure}

The combined optical pump and absorption layer is used for increasing the on-chip optical excitation efficiency. In the pumping diode, charge carriers are injected into the InGaAs QW when applying a forward bias across the p-i-n junction. Carrier recombination in this 2D layer gives rise to broadband optical emission around 980nm as can be seen in the blue spectral curve displayed in Figure \ref{fig:OCP}(e), measured in a confocal microscope setup from the top, on a position with no QD on the pumping diode. The spectral signal looks rather weak as this light is non-resonant with the embedded vertical cavity which favours in-plane rather than out-of-plane emission. Pumping light is absorbed by resonant optical excitation of confined states in the QW of the tuning diode. Here, carriers can recombine in QDs resulting in emission of single or entangled photons in the telecom O-band around 1310nm along the vertical direction as these are resonant with the embedded cavity. A corresponding spectrum is displayed as red curve in Figure \ref{fig:OCP}(e). The geometric and extreme spectral separation of pumping light and QD emission of more than 300nm enables an excellent suppression of background using standard optical filtering methods, resulting in high-purity single-photon emission (a measurement of a second order autocorrelation function is shown in Figure \ref{fig:cor}\,(a)). In comparison with devices where the on-chip optical excitation is only based on conventional wetting layer emission and absorption \cite{lee2017electrically} (thickness $< 1$nm), we observe a significantly enhanced efficiency of the optical pumping process. This is most likely caused by increased absorption efficiency of pumping light due to the much higher thickness (5nm) of the pump and absorption layer.

\begin{figure}[!th]
\centering
\includegraphics[width=0.55\textwidth]{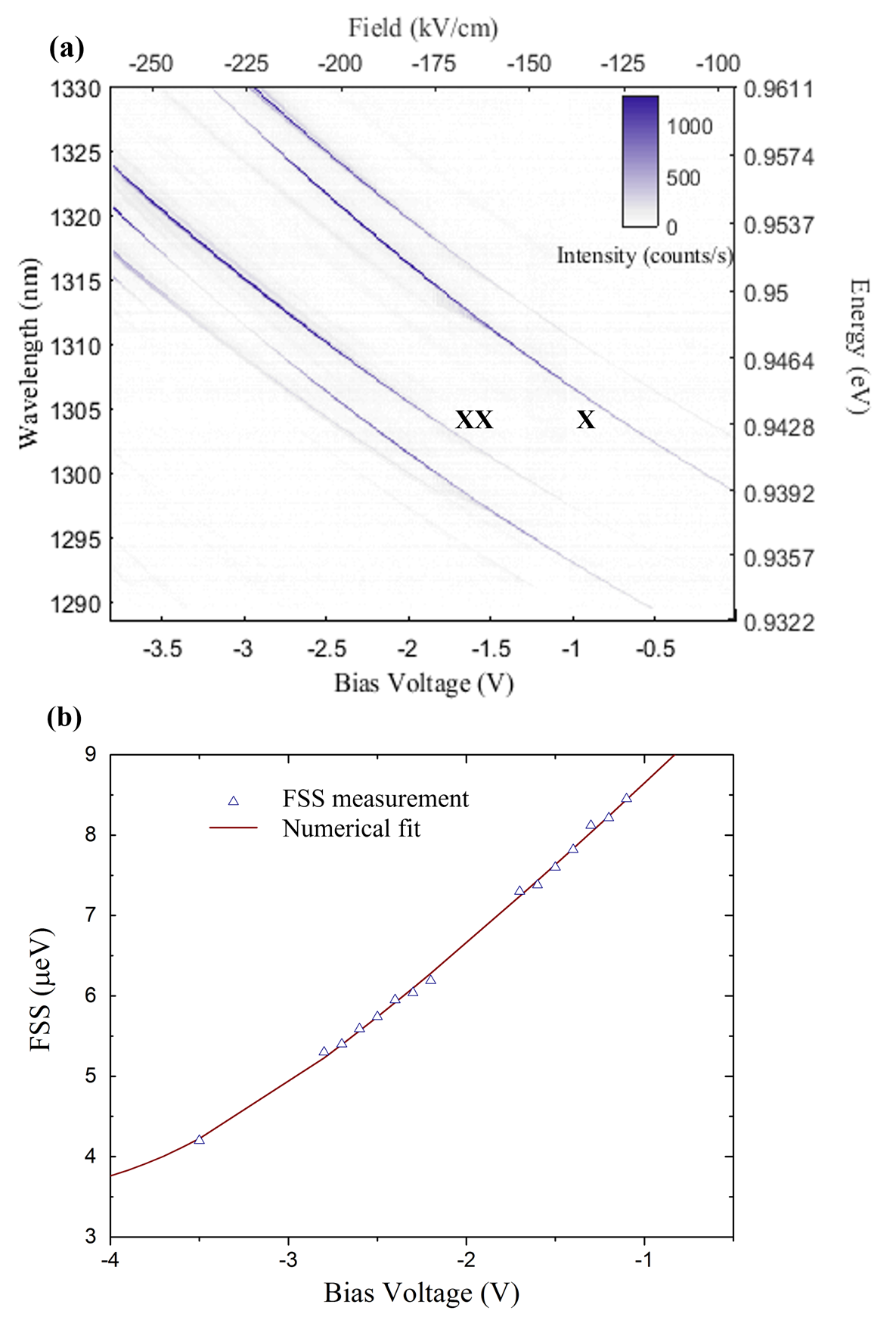}
        \caption{Tuneability of photon emission. (a) Emission spectrum of on-chip optically excited single QD as a function of applied bias voltage to the tuning diode. A wavelength shift of around 30nm is observed for XX and X emission. (b) Corresponding change of the fine structure splitting (FSS) of the QD as a function of the applied bias. The splitting parameter was extracted from the time-resolved beat of X-XX correlations when filtering with a linear polarizer. Owing to the high precision of this method, error bars are negligible. The red curve is a numerical fit based on the theoretical model for dependence of the splitting due to the quantum confined Stark effect \cite{pooley2014energy}.}
         \label{fig:tun}
\end{figure}

The AlGaAs barriers surrounding the QD layer as shown in Figure \ref{fig:OCP}(d) inhibit escape of charge carriers, enabling the application of large electrical fields across the \textit{p-i-n} junction. Growth of the QDs in the InGaAs QW further loosens the confinement of electron-hole pairs, resulting in a higher permanent dipole moment and polarizability of excitons, enhancing the tuneability of QD emission via the quantum confined Stark effect \cite{patel2010two, miller1984band, laucht2009electrical}. In the following, we selected a single QD with good entanglement fidelity and intensity of emitted biexciton (XX) and exciton (X) photons. Figure \ref{fig:tun}(a) shows wavelength tuneability of its emission spectrum when driving the pumping diode with a forward bias and changing the reverse bias on the tuning diode from -3.8V to 0V. The neutral exciton (X) and biexciton (XX) states can both be shifted over the center of the telecom O-band (1310nm) with a maximum range of more than 25nm for the applied voltages. This significantly exceeds the tuneability performance of short-wavelength devices \cite{patel2010two, lee2017electrically} and meets the requirements for compatibility with the ITU CWDM grid. It has to be emphasized that the electrical driving conditions for excitation are comparable to directly electrically injected ELEDs \cite{salter2010entangled}  indicating the optimized operation of the tuneable devices.

We also observe the characteristic tuning of the fine structure splitting (FSS) of neutral X and XX states as a function of bias voltage, as shown in Figure \ref{fig:tun}(b). For entangled photon-pair emission it is important to achieve FSS values of 10\(\mu eV\) or less such that indistinguishability of the resulting two spectral components of a photon is guaranteed with typical detection setups featuring 100ps timing resolution or better. The splitting of the selected QD stays significantly below that threshold over the entire tuning range, enabling the demonstration of a tuneable entangled photon pair source. In the remainder of the paper, we operate the device with a voltage of -2.6V applied to the tuning diode, resulting in a FSS of 5.6\(\mu eV\) and XX emission at 1310.00nm (the exact center of the O-band) and X emission at 1321.45nm.

\begin{figure}[!th]
\centering
    \includegraphics[width=\textwidth]{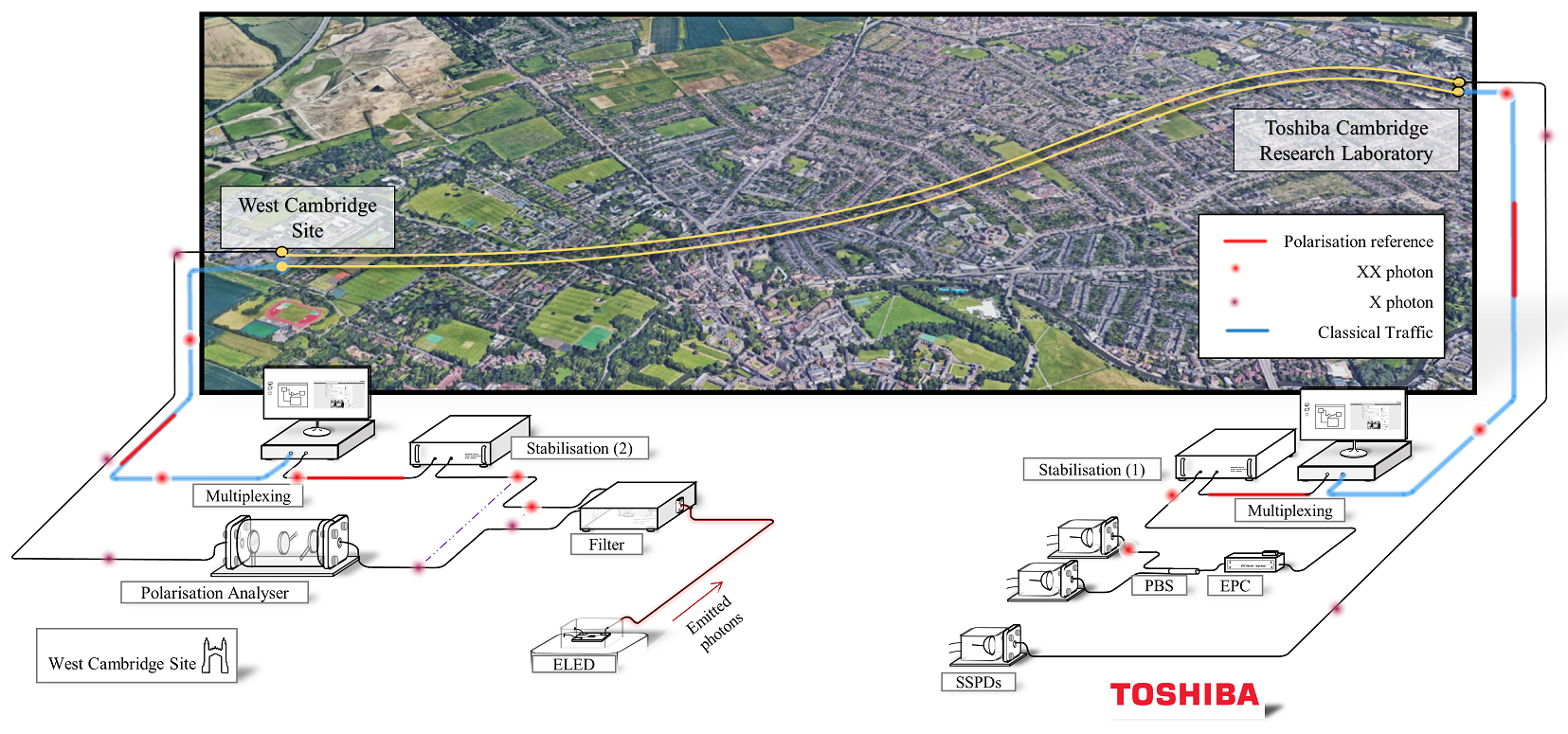}
  \caption{Schematic of overall experimental setup for entangled photon transmission from network-deployed ELED. Optical signals from 3 sub-systems are sent over the network. Single photons are shown as ‘red’ (XX) and ‘violet’ (X) dots, laser pulses for polarisation stabilisation are shown as short ‘red’ lines and classical data traffic is shown as continuous ‘blue’ line. Entangled photon pairs are generated from the tuneable ELED device at West Cambridge (CAM). X and XX photons are spatially separated with a spectral filter module. X photons pass through a polarisation analyser consisting of a half waveplate (HWP), a quarter waveplate (QWP) and a linear polariser (LP) before being sent over one of the two installed fibers. Entangled XX photons are sent over the other field fiber which is stabilised for polarisation drifts using a polarisation stabilisation sub-system (Stabilisation 1 and 2). At CRL, their quantum state is measured using an electronic polarisation controller (EPC), a polarising beam splitter (PBS) and two superconducting single photon detectors (SSPDs). For remote control of the deployed source and system components, a classical data connection in the telecom C-band is established over the same fiber using a classical-quantum multiplexing system (labelled as Multiplexing). Arrival time of the X photons in the first fiber is measured with a third SSPD at CRL. \newline Map: Imagery \textcopyright 2019 Google, Getmapping plc, Infoterra Ltd \& Bluesky, The GeoInformation Group, Maxar Technologies, Data SIO, NOAA, U.S. Navy, NGA, GEBCO, Maxar Technologies, Landsat / Copernicus, Map data \textcopyright 2019}
\label{fig:setup}
\end{figure}

Simple electrical operation and compatibility with wavelength division multiplexing standards allow us for the first time to demonstrate the deployment and telecommunication network integration of the device. We operate a fully remotely controlled system that is installed in an office at a location in West Cambridge (CAM), which is connected via the Cambridge Network with two optical fibers of 15km length each to the Toshiba Cambridge Research Laboratory (CRL) in the Science Park. Figure \ref{fig:setup} shows the overall experimental setup and a map of Cambridge. The system consists of multiple sub-systems: an entangled photon transmission module containing the tuneable ELED, a spectral filter with polarization reference for detector calibration, a polarisation stabilisation module and a quantum/classical data multiplexing module.

The ELED device is mounted in a closed-cycle cryostat, cooling it to 6K. The emitted light from the quantum dot is coupled into single mode optical fiber and sent to a spectral filter module mounted in an instrument rack. Here, a diffraction grating is used to separate the entangled X and XX photons into two different output modes. Right after the spectral filter, the X photons pass through a polarisation discrimination setup that consists of a half waveplate (HWP), a quarter waveplate (QWP) and a linear polariser (LP), projecting their quantum state into different polarisation bases at the remote location before transmission to CRL for detection. The entangled XX partner photons are sent to CRL over a separate network fiber which is stabilised for changes of birefringence \cite{xiang2019long} and additionally hosts a classical data connection for remote control of the deployed system. At CRL, their polarisation state is projected using an electronic polarisation controller (EPC) and a fibre polarising beam splitter (PBS). Single-photon arrival times for X and XX photons are recorded using superconducting nanowire single photon detectors and photon-pair correlations in different detection bases are measured for evaluation of the entanglement fidelity across the 15km fiber link.

A precise alignment of the detection basis to the eigenbasis of the QD emission is essential in these experiments. In this work, we make use of a new method for the exact measurement of said eigenbasis, that is based on analysing time-resolved photon-pair correlations in a set of randomly oriented detection bases (see appendix). For detector calibration, precisely controlled polarisation reference light is injected into X and XX output modes from the spectral filter module.

\begin{figure}[!t]
\centering
\includegraphics[width=0.5\textwidth]{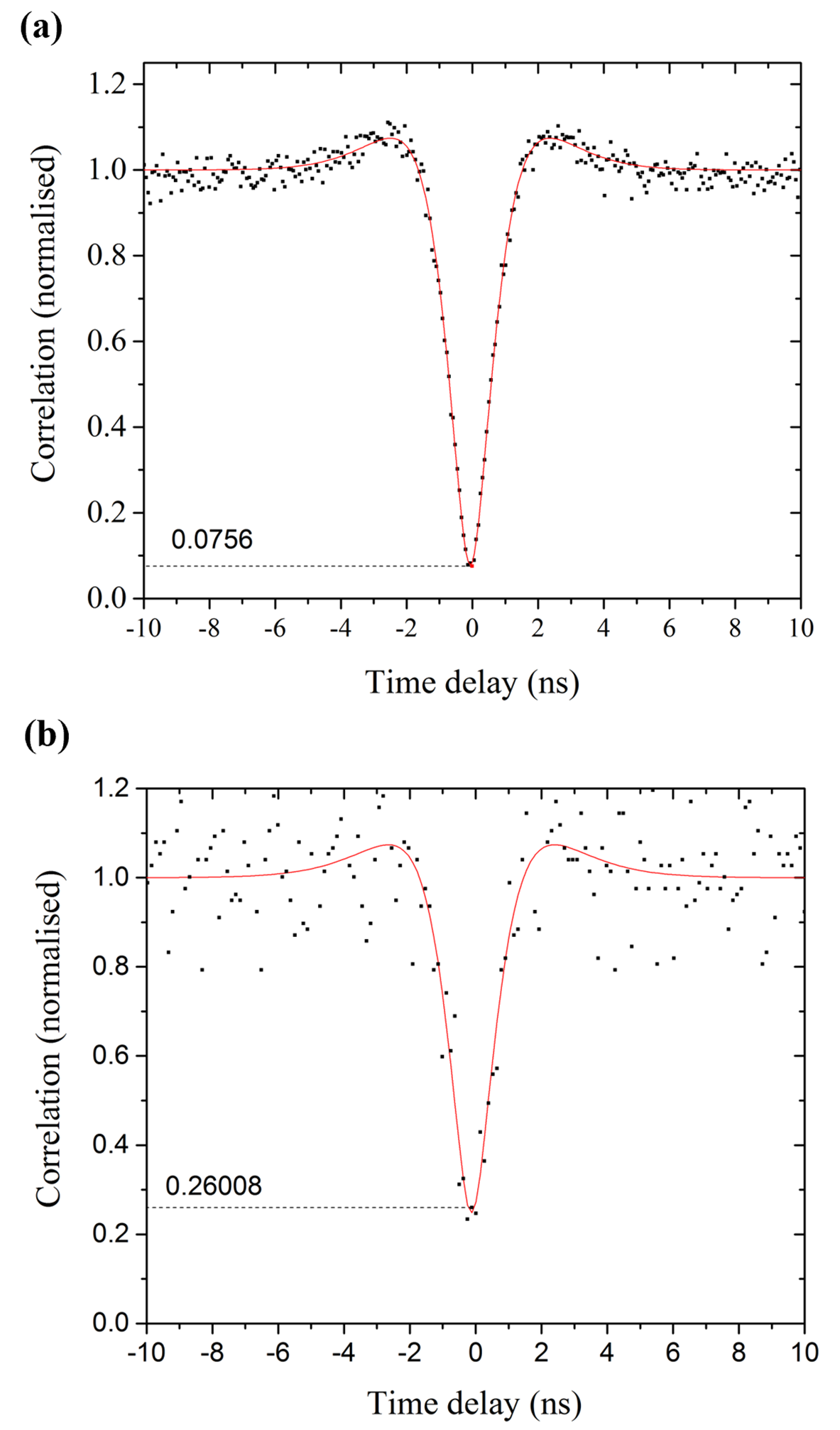}
        \caption{Second-order autocorrelation measurement result for XX photons (a) in the laboratory and (b) after transmission over 15km of installed fiber, multiplexed with classical data traffic. Data acquisition times were 5 minutes and 30 minutes respectively. The red curves show a theoretical fitting function which is used to extract the displayed \(g^{(2)}(0)\) values and corresponding background contribution.}
        \label{fig:g2}
\end{figure}

For demonstrating true network integration, the entangled quantum bits are multiplexed with classical communication traffic over the same optical fiber using coarse wavelength division multiplexing with quantum light at 1310nm and a bidirectional 1Gbit data connection at 1550nm. The link is established using standard C-band transceiver modules with the launch power set to 7.3$\mu$W and up- and down-stream traffic is isolated by directional filtering, using circulators. Multiple CWDM modules in the quantum and classical channels and a spectral filter with a FWHM of less than 1nm installed in the quantum channel at CRL, are used to suppress the background from the classical light to a very low level. Previous field trials have illustrated the feasibility of multiplexing QKD qubits based on attenuated laser pulses with classical data traffic over installed network links \cite{townsend1997simultaneous, mao2018integrating, choi2014field} . However, the co-existence of entangled qubits from a sub-Poissonian photon source and classical data traffic over a real-world network has not been demonstrated yet. The classical communication channel is used to remotely control the ELED system, the polarisation stabilisation system and data acquisition during the experiments.

Figure \ref{fig:g2} shows a comparison of the single photon purity of quantum light from the ELED before and after transmission over the lit field fiber, by measuring the second order autocorrelation function. After transmission, the light is still strongly anti-bunched with a g(2)(0) value of 0.26 being significantly below the classical limit of 0.5. A more detailed analysis taking signal and background contributions for both measurements into account reveals that the increase from the laboratory-measured value of 0.076 is primarily caused by a reduction of the signal-to-background ratio due to photon loss in the network fiber and all system components (11.8dB in total), and not the presence of the classical data link.

Entanglement measurements are carried out by recording photon-pair correlations \(c_{PQ}\) for co- and cross-polarised configuration in the three principal detection bases horizontal/vertical (HV), diagonal/anti-diagonal (DA) and right-/left circular (RL) when projecting the quantum states of X photons at CAM and XX photons at CRL. Figure \ref{fig:cor} (a)-(c) shows the corresponding results. The detection basis is changed every ten minutes. In the HV basis, we observe a strong correlation of photon polarisations with a maximum contrast of 96.8\%. The superposition bases DA and RL show high contrast quantum correlations as well, with the characteristic beat owing to the fine structure splitting of the QD as reported multiple times before  \cite{ward2014coherent, stevenson2008evolution, akopian2006entangled}. The fidelity to the maximally entangled Bell \(\phi^+\)state is then calculated as \(F = (1+C_{HV}+C_{DA}-C_{RL})/4\) where \(C_{PQ}\) are the so-called correlation coefficients with \(C_{PQ}={(c}_{PP}-c_{PQ})/{(c}_{PP}+c_{PQ})\). The fidelity peaks at \(95.19\pm 0.5 \%\) (see Figure \ref{fig:cor}(c)) for a post-selection window size of 48ps. To demonstrate the good stability of the deployed ELED device, the experiment was continuously running over a period of 40 hours. Figure \ref{fig:cor}(e) shows the evolution of the entanglement fidelity based on 120 minutes of data per displayed point. The bottom of the graph shows the corresponding quantum-bit error rate which is stable around 3.4\%. These results are directly obtained from raw-data and no background subtraction of any kind has been done during processing. The classical communication link was running at all times.

To conclude, we have developed the first electrically operated tuneable ELED emitting in the main telecommunication wavelength band suitable for multiplexing with classical communication signals. Combination of on-chip optical excitation with dedicated layer structure and device design enable low-voltage operation (\(<5\)V) and large wavelength tuneability (\(>25\)nm). A high single photon purity compatible with typical non-resonantly excited QD light sources was observed and entanglement with a fidelity of more than 95\% was achieved after integration and deployment in a real-world fiber network. Continuous operation of the source showed stable entanglement over 15km of installed fiber shared with classical data traffic used for remote control of all system components.

Compatibility of the device with standard low-voltage power supplies and compliance with laser-safety standards due to exclusively electrical operation pave the way for future integration in various end user applications. The large wavelength tuneability furthermore opens up the route for flexible wavelength division multiplexing of multiple tuneable ELED sources over the same optical network link and most importantly, multiplexing with classical data channels. This is of fundamental importance for the low-cost integration of quantum-networks in classical network infrastructures.

\section*{Funding Information}
The authors acknowledge partial financial support from the Engineering and Physical Sciences Research Council, and the EPSRC Quantum Technology Hubs in Quantum Communications. ZX would like to show his gratitude to the Cambridge Trust, China Scholarship Council and Toshiba Research Europe Limited for the financial support of his Ph.D. programme.

\section*{Acknowledgements}
The authors would like to thank A. J. Bennett for a critical reading of the manuscript, J. Dynes for technical discussions and A. Wonfor for arrangement of the access to the fiber network.


\section*{Appendix}

\subsection*{Calibration of detection basis}

The evaluation of the entanglement fidelity from the QD source is typically done by correlation measurements in the HV, DA and RL bases in the reference frame of the eigenbasis of QD emission. Therefore, a precise alignment of the measurement setup to the eigenbasis of the QD emitter is essential. Conventional alignment methods make use of a calibrated reference that is injected into the main optical path towards the detectors, before the entangled photons are being coupled into the first optical fiber \cite{young2006improved} as birefringence would otherwise rotate the linear (HV) eigenstates into unknown elliptical states. However, this configuration is usually bulky and not compatible with a compact collection setup as in this work, which only consists of a single collection lens (NA=0.5) and a fiber collimator outside of the cryostat window. We therefore developed a precise technique for detection of the QD eigenbasis after photons have propagated through birefringent media such as single mode fiber. This enables the injection of the correct polarisation references for detector calibration at an arbitrary point between the source and detectors.

The method works as follows. First we take X-XX correlation measurements (co- and cross-polarised) in three well-known arbitrary but linearly independent detection bases in the reference frame of the polarisation reference. As will be explained in the following paragraphs, the time-dependent evolution of the correlation signal is then used to extract the exact orientation of each of the three reference states in the reference frame of the eigenbasis of QD emission. Using the standard procedure for Müller matrix evaluation, this allows us to calculate the transformation matrix between polarisation reference and QD eigenbasis which enables the generation of perfectly matching reference states for detector calibration in the second step.

\begin{figure}[!th]
\centering
    \includegraphics[width=0.87\textwidth]{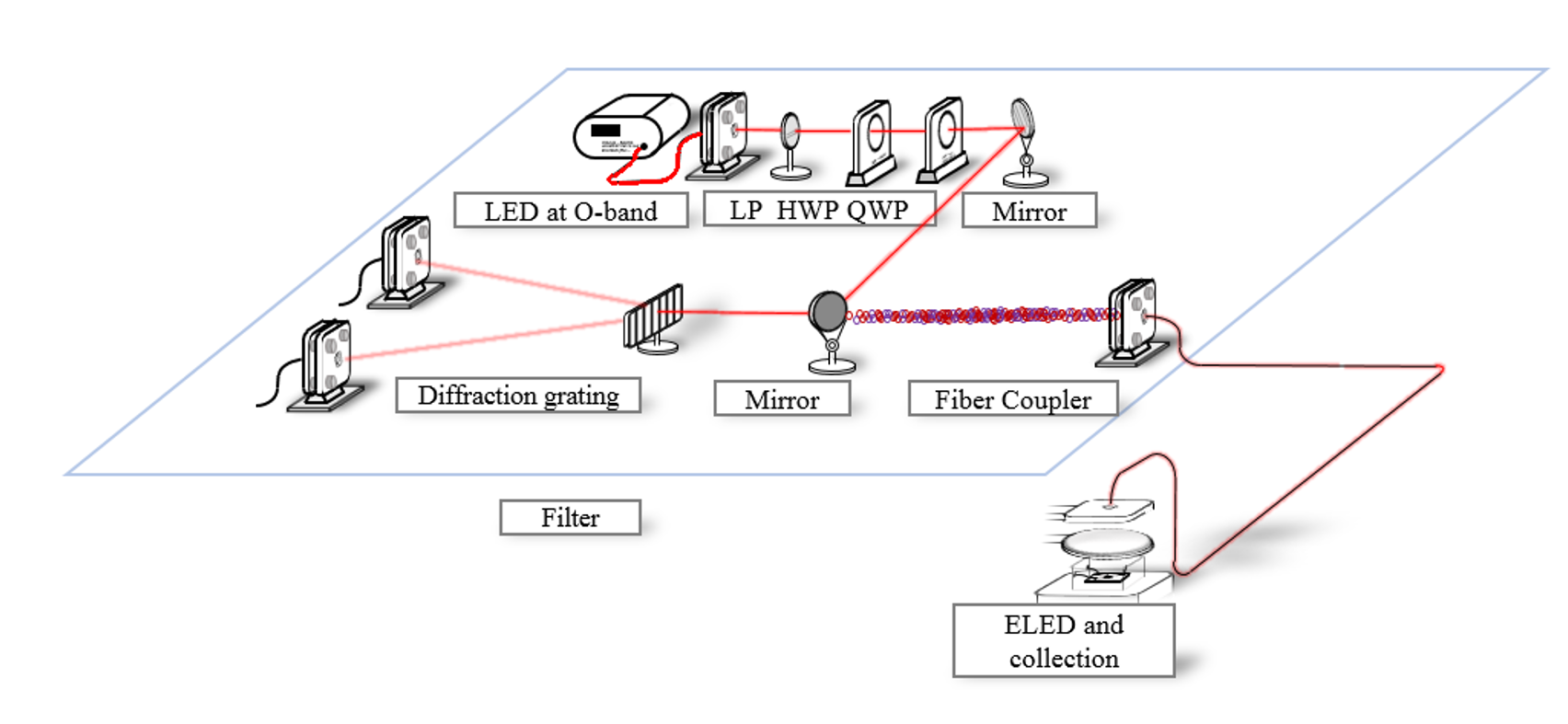}
  \caption{Experimental setup for injection of polarisation reference to qubit measurement system. A broadband LED emitting in the telecom O-band passes a LP, a HWP and a QWP to generate arbitrary controllable polarisation reference states at X and XX wavelength.}
\label{fig:ali}
\end{figure}

The experimental setup is shown in Figure \ref{fig:ali}, corresponding to the spectral filter module shown in Figure  \ref{fig:setup} in the main text. Entangled photons from the ELED are coupled into single mode fiber which guides them to the module which is mounted in a separate instrument rack a few meters apart. A flip mirror is installed just before the diffraction grating in the filter, which is used to switch between the transmission of entangled photon pairs from the ELED and injection of a polarisation reference beam into the main optical path.

For a QD with fine structure splitting \(\delta\), the emitted entangled state has a time-dependent phase variation that can be described by the following equation:

\begin{equation}
\left|\varphi\right\rangle=\left|H_X\right\rangle\left|H_{XX}\right\rangle+e^{i\frac{\delta}{\hbar}t}\left|V_X\right\rangle\left|V_{XX}\right\rangle
\end{equation}
Where \(\left|P_{X}\right\rangle \) and \(\left|P_{XX}\right\rangle \) represents polarisation state \(\left|P\right\rangle\) in the eigenbasis of QD emission for X and XX photon, respectively. Let us assume that the detection system is aligned for projection to a randomly oriented polarisation state \(\left|H^{\prime}\right\rangle\) described by the parameters \(\theta\) and \(\varphi\) in spherical coordinates on the Poincaré sphere:
\begin{equation}
\left|H^{\prime}\right\rangle=\cos{\frac{\theta}{2}}\left|H\right\rangle+\sin{\frac{\theta}{2}}e^{i\varphi}\left|V\right\rangle
\end{equation}
Projection of XX photons to \( \left|H^{\prime}\right\rangle \) will collapse the exciton photon into state
\begin{equation}
\left|{P^{\prime}}_X\right\rangle=\left\langle{H^{\prime}}_{XX}|\varphi\right\rangle=\cos{\frac{\theta}{2}}\left|H_X\right\rangle+\sin{\frac{\theta}{2}}e^{i(\frac{\delta}{\hbar}t-\varphi)}\left|V_X\right\rangle\ .
\end{equation}
The probability for the detection of a co-polarised coincidence between X and XX photons is then described by
\begin{equation}
\label{eq:pco}
p_{co}={\left|\left\langle {P^{\prime}}_X|H^{\prime}\right\rangle\right|}^2=1-2\sin^2{\frac{\theta}{2}}\cos^2{\frac{\theta}{2}}\left(1-\cos{\left(\frac{\delta}{\hbar}t-2\varphi\right)}\right)\ .
\end{equation}

\begin{figure}[!b]
\centering
    \includegraphics[width=0.55\textwidth]{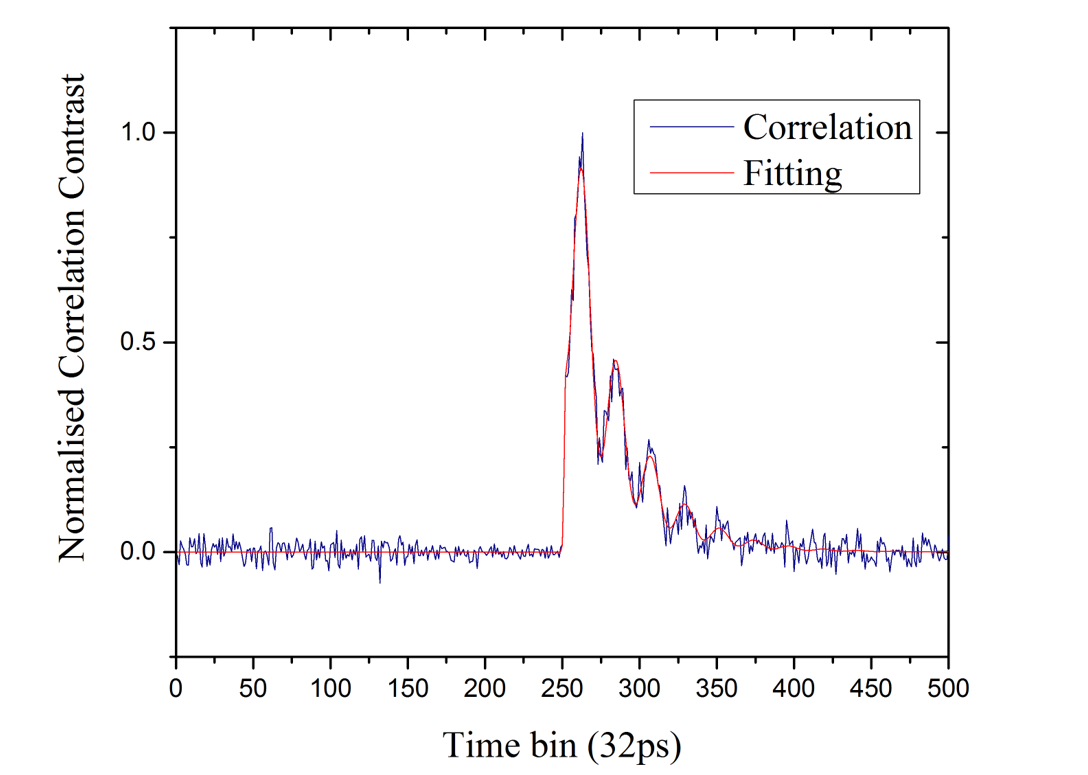}
  \caption{Normalised difference of co- and cross-polarised photon coincidences measured in a random detection basis. The red curve shows an experimental fit according to equation (\ref{eq:dpq}) for extraction of the fine structure splitting and angular information of the detector basis orientation.}
\label{fig:fit}
\end{figure}

Experimentally, we measure co- and cross-polarised photon coincidences \(c_{PP}\) and \(c_{PQ}\) and calculate the normalised difference \(d_{PQ}=A\left(c_{PP}-c_{PQ}\right)\) with $A$ being a normalization constant (see Figure  \ref{fig:fit}). This relates then to equation (\ref{eq:pco}) as
\begin{equation}
\label{eq:dpq}
d_{PQ} =\left(2 p_{co} - 1\right) e^{-\frac{t}{\tau}}
\end{equation}
where \(\tau\) is a constant accounting for exponential decay of the photon correlation due to the natural lifetime of the X state and other spin dephasing processes including repumping. A numerical fit to experimental two-photon correlations in a certain detection basis therefore enables the extraction of precise values for the fine structure splitting \(\delta\) and for angles \(\theta\) and \(\varphi\). The relation between these angles and the corresponding Stokes parameters \(s_1,\ s_2,\ s_3\) is given by
\begin{equation}
s_1\ =\ cos\ \theta\ \ \  \\
{\ \ \ \ s}_2\ =\ sin\ \theta\ cos\ \varphi \\
{\ \ \ \ s}_3\ =\ sin\ \theta\ sin\ \varphi\ .
\end{equation}

The polarisation transformation of an optical fiber or other linear optical elements is conveniently described by the Müller matrix $M$. 
\begin{equation}
S_m\ =\ M\ S_r
\end{equation}
Where \(S_r\) is the Stokes vector for the orientation of the polarisation detection system, being calibrated by a corresponding reference state. \(S_m\) is the Stokes vector extracted from fitting equation (\ref{eq:dpq}) to photon pair correlations, which describes the orientation of the QD eigenbasis with respect to the calibration reference state. It is straight forward to reconstruct $M$ from a set of calibration states (e.g. \(S_r\in\left\{H,D,R\right\}\)) and their corresponding measured vectors \(S_m\) \cite{layden2012optimum}. Knowledge of $M$ then enables deterministic setting of reference calibration states \(S_{r\prime}\) that perfectly match the principal polarisation states \(S_{eigen}\) in the reference frame of the QD emitter.

\begin{equation}
S_{r\prime}\ =\ M^{-1}\ S_{eigen}\ .
\end{equation}

\end{document}